\documentclass[conference]{IEEEtran}
\IEEEoverridecommandlockouts
\usepackage{cite}
\usepackage{amsmath,amssymb,amsfonts}
\usepackage{algorithmic}
\usepackage{graphicx}
\usepackage{textcomp}
\usepackage{xcolor}
\usepackage{url}
\usepackage{todonotes}
\usepackage{comment}

\usepackage{booktabs}
\usepackage{multicol}
\usepackage{multirow}

\def\BibTeX{{\rm B\kern-.05em{\sc i\kern-.025em b}\kern-.08em
    T\kern-.1667em\lower.7ex\hbox{E}\kern-.125emX}}

\begin{document}

\title{Heart Rate Detection Using an Event Camera\\
\thanks{This research was conducted with partial financial support of Science Foundation Ireland [12/RC/2289\_P2] at Insight the SFI Research Centre for Data Analytics at DCU..}
}

\author{
            \IEEEauthorblockN{Aniket Jagtap,\\
RamaKrishna Venkatesh Saripalli}
        \IEEEauthorblockA{\textit{School of Computing} \\
                            \textit{Dublin City University}\\
                            Glasnevin, Dublin 9, Ireland}
\and
        \IEEEauthorblockN{Joe Lemley}
        \IEEEauthorblockA{\textit{Xperi Corporation} \\
                            \textit{Parkmore Indl. Estate}\\
                            Galway, Ireland}
                         \and
\IEEEauthorblockN{Waseem Shariff}
        \IEEEauthorblockA{\textit{University of Galway}\\
                            Galway, Ireland}
                            \and
        \IEEEauthorblockN{Alan F. Smeaton}
        \IEEEauthorblockA{\textit{Insight Centre for Data Analytics} \\
                        \textit{Dublin City University}\\
                        Glasnevin, Dublin 9, Ireland \\
                        alan.smeaton@DCU.ie}

}

\maketitle

\begin{abstract}
Event cameras, also known as neuromorphic cameras, are an emerging technology that offer  advantages over traditional shutter and frame-based cameras, including high temporal resolution, low power consumption, and selective data acquisition.
In this study, we propose to harnesses the capabilities of event-based cameras to capture subtle changes in the surface of the skin caused by the pulsatile flow of blood in the wrist region. We investigate whether an event camera could be used for continuous noninvasive monitoring of heart rate (HR). Event camera video data from  25  participants, comprising varying age groups and skin colours, was collected and analysed.  Ground-truth HR measurements obtained using conventional methods  were used to evaluate of the accuracy of automatic detection of HR from event camera data.
Our experimental results and comparison to the performance of other noncontact HR measurement methods  demonstrate the feasibility of using event cameras for pulse detection.
We  also acknowledge the challenges and limitations of our method, such as light-induced flickering  and the sub-conscious but naturally-occurring tremors of an individual during data capture. 
\end{abstract}


\section{Introduction}

In recent years event cameras have emerged as a novel imaging paradigm and an alternative to  conventional shutter-based or frame-based cameras. The potential for event camera applications span a wide array of industries including robotics or wearable electronics, where fast latency, reduced power consumption, and functioning in unpredictable lighting conditions are crucial \cite{gallego_event-based_2022}.

Traditional shutter-based cameras acquired video content by opening and shutting a physical shutter at specified intervals. These have been replaced by conventional  frame-based cameras where light coming through the camera lens reaches a light sensor where
photovoltaic conversion of light to electrical signals happens synchronously at up to millions of photosites. Each of these photosites corresponds to a pixel in the resulting image and in the case of video, this simultaneous conversion of light to electrical signals happens at fixed intervals, normally 25 or 30 times per second.

Due to their fixed interval approach, conventional frame-based cameras encounter a challenge known as ``undersampling". This phenomenon leads to information loss when attempting to capture events at the microsecond level \cite{chen2020event}. However, a promising alternative emerges with neuromorphic event cameras. Event cameras are a type of imaging sensor that responds to local changes within their field of view. Unlike traditional cameras, event cameras record pixel-level brightness asynchronously and independently. They do so in response to alterations in scene luminance, rather than adhering to predetermined frame intervals. Data recorded in an event camera is made up of a stream of information packets, each with the $x$ and $y$ coordinates or pixel locations, a timestamp for the recording, and an indication of the brightness change which caused the information packet to be generated.  The temporal resolution of an event camera is fine-grained where light sensors can record a light change and generate a packet at microsecond level. This capability not only allows event cameras to bypass the issues of ``undersampling" and motion blur but also enables them to achieve real-time detection of rapid luminance changes. Events are recorded with accuracy down to the microsecond, and they can achieve equivalent frame rates surpassing 10,000 frames per second \cite{kielty2023neuromorphic}.

The inherent attributes of event cameras, encompassing their remarkable temporal precision, minimal time lag, and extensive dynamic range, hold significant promise for enabling accurate, real-time, and non-contact monitoring of a driver's heart rate (HR). This study is a proof-of-concept (PoC) dedicated to delving into the potential applications of event cameras in this domain. To achieve this we engage in a series of data collection activities to gather event camera data from a cohort of 25 participants. Concurrently, we collect ground truth heart rate measurements through the utilisation of smartwatches worn by the subjects. This dual-source data acquisition strategy equips us with a comprehensive dataset. Following the data collection phase, we begin our analysis. We center our attention on identifying shifts in event camera polarity within the region of interest, specifically around the wrist area. By carefully processing, we can estimate the underlying heart rates of the observed individuals. These estimated heart rates are subsequently compared against the  ground truth measurements acquired from a smartwatch. Using the quantitative metrics such as mean average error (MAE) and root mean squared error (RMSE) we validate the proposed approach. 

\section{Background}

\subsection{Periodicity in Data}
\label{sec:periodogram}

Periodicity is a property of a time series of data whereby  a pattern  recurs within a data stream at regular or periodic intervals. This basically refers to the regularity of things that occur repeatedly in nature as typical behaviour which can be captured in data. Deviations from regular or periodicity data  are referred to as outliers. The concept of periodicity is used in complex systems to discover insights within the patterns which can lead to deeper understanding of the data and the underlying natural phenomenon \cite{2023} and the distribution of frequencies in a data stream such as from an event camera, is called a periodogram.

One example of periodicity occurring in natural systems is heart rate or the number of beats of a heart in a given period, typically 1 minute. The human (and other animal) heart beats with a regular frequency which changes only slowly. When we are at rest, sitting for example, it may beat at 70 beats per minute and when we get up to walk somewhere it may rise to perhaps 100 beats per minute but this will happen gradually, not instantly.  The human HR or pulse can be detected using a  device which picks up the electrical signals within the body which control the heart beating through contact sensors placed on the skin. This approach is used in medical devices such as an electrocardiogram and has recently started to appear in consumer devices such as the Apple watch.

\subsection{Measuring Heart Rate}

A common approach to measuring HR uses photoplethysmography which is based on using green LED lights that flash on and off with high frequency and when paired with light-sensitive photodiodes they detect the flow of blood through the wrist on a continuous basis.  As a result they can detect the pulsation of blood flow which is in sync with the heart beating and from this can determine HR. This approach is also popular on consumer devices such as the Apple watch.  

There are several noncontact pulse rate measurement systems based on the technique of photoplethysmography which measure both HR and heart rate variability (HRV) using an optical camera and which compensate for subject movement e.g. \cite{BERES2021102589,s17071490}. This has an acceptable mean absolute error and root mean square error (MAE/RMSE) of 2.11/2.93, 2.43/3.44, and 2.26/3.45 beats per minute (bpm) for biking, stepping, and treadmill exercises, respectively \cite{9654212}.

The human heart rate may also be determined manually by sensing the motion at parts of the body where an artery is close to the skin, such as the radial artery in the wrist the carotid artery in the neck or the near the superficial temporal artery near the temple on the head. It can also be measured in any place that allows an artery to be compressed near the surface of the body, such as at the neck, groin, behind the knee, and near the ankle joint.  For some of these areas, notably the wrist and the neck, the pulsation from the artery can sometimes be observed on the surface of the skin as a throbbing motion, whose periodicity is the heart rate of the subject.  Mostly however, this throbbing movement is so minor that it is not visible to the naked eye.

With respect to our use of an event camera, in this work we set out to use the concept of periodicity for determining pulse rate of an individual from observations of the movement on the inner surface of the wrist caused by the radial artery, similar to  work reported in \cite{rong2021radar} though that work used very low power, non-ionizing radio frequency signals whereas we set out  out to determine if an event camera can be used to detect pulse rate in humans from the sometimes invisible tiny movements which happen on the surface of the skin caused by  radial artery pulsation in the wrist. We  gather and detect event camera events from the wrist areas of a set of subjects and  use periodicity detection algorithms \cite{2023} on these events to see if we can identify a recurring pattern which has the same periodicity as the pulse rate of the subject.
We now proceed to describe our experimental design in the next section.

\subsection{Event Cameras in Biomedical Applications}

To date, the application of event cameras in biomedical contexts remains largely unexplored, offering a realm of untapped possibilities. Presently, there are no established studies that have delved into the utilisation of event cameras for specific biomedical purposes. However, the concept itself stands as a promising proof of concept, suggesting that event cameras possess attributes conducive to innovative applications in this domain.

Event cameras, renowned for their high dynamic range (HDR) and remarkable temporal resolution, emerge as a compelling imaging technology well-suited to scenarios necessitating the capture of rapid motion. Their inherent sensitivity to light enhances their utility in low-light environments, positioning them as a preferable choice for applications like driver monitoring. Notably, event cameras can achieve an extraordinary dynamic range of up to 140~dB, a considerable advancement over conventional frame-based cameras that typically offer around 60~dB \cite{holesovsky_experimental_2021}.

These attributes collectively render event cameras as prospective contenders with substantial potential to effectively match the high-frequency sensing requirements characteristic of physiological parameters such as heart rate. Despite the dearth of pertinent studies to date, the fundamental attributes of event cameras as high-speed, high dynamic range sensors position them as intriguing candidates for future exploration in the domain of physiological sensing and monitoring.

\section{Methodology and Data Gathering}

\begin{figure}[!htb]
    \centering
    \includegraphics[width=\linewidth]{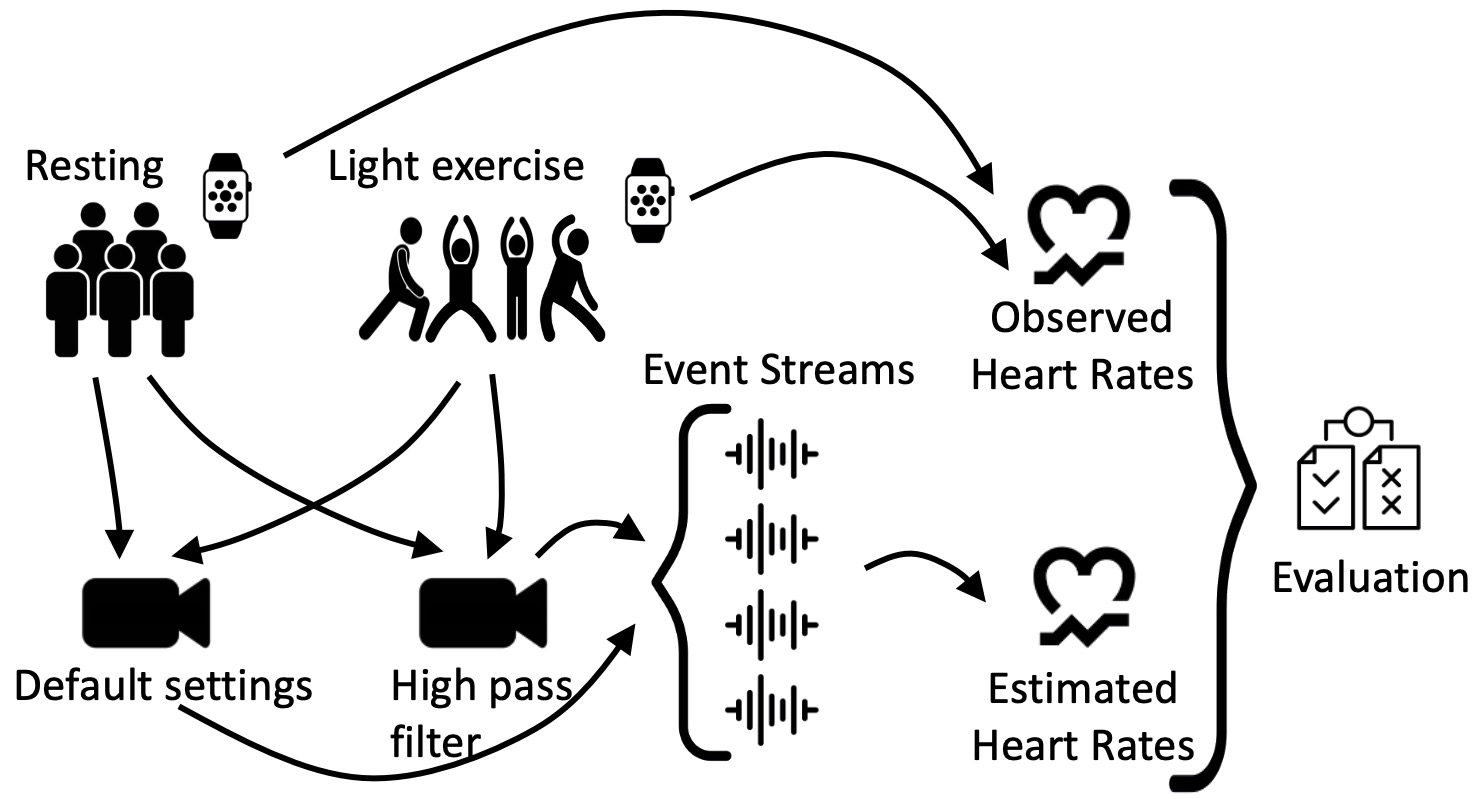}
    \caption{Methodology for our experiments.}
    \label{fig:data-flow}
\end{figure}

Figure~\ref{fig:data-flow} provides an overview of our experimental pipeline. In this study, an event camera is tested with various bias settings. Using this sensor, pulse rates are measured in two setups: subjects at rest and subjects after completing some light exercise. Event camera recordings and smartwatch-based ground truth heart rates are collected in both setups. The accuracy of predicted heart rates from the event camera is evaluated against observed heart rates from the smartwatches.

\subsection{Camera Configuration and Bias Settings}
A Prophesee EVK4 event camera\footnote{Prophesee, Paris, France \url{https://www.prophesee.ai/event-camera-evk4/}} was used to gather event stream recordings for the  experiments in this paper. Camera manufacturers introduce biases to enable users to have some degree of freedom to enhance control over the camera's output. These biases allow adjustments like altering sensor sensitivity to light variations, managing the event generation count, and performing similar operations that affect sensor-level changes in camera configurations.  The sensor performance of can be adjusted for a variety of application requirements and environmental situations yielding faster speed, lower background activity, higher contrast sensitivity threshold, and more. Below are  key bias settings along with their functionalities \cite{prophesee_doccumentation}:

\begin{enumerate}
    \item bias\_diff\_on which adjusts the contrast threshold for ON events. This determines the ON contrast threshold, the factor by which a pixel must get brighter before an ON event occurs for that pixel. It is usually adjusted when the user wants to change how many events are output during a big change in illumination, or else to change the sensitivity to small positive light changes.
    \item bias\_diff\_off which adjusts the contrast threshold for OFF events. This determines the factor by which a pixel must get darker before an OFF event occurs for that pixel. It is usually adjusted when the user wants to change how many events are output during a big change in illumination, or else to change the sensitivity to small negative light changes.
    \item bias\_fo  adjusts the low-pass filter which changes how rapidly fluctuating light is filtered out. It determines the maximum rate of change of illumination that can be detected and is often used to remove flickering and noise in a scene, but doing so will also increase the latency of the sensor.
    \item bias\_hpf  adjusts the high-pass filter which determines how slow changes in illumination  are filtered out. It determines the minimum rate by which the light must change for a pixel to output an event. It is often used to remove the background of a scene and to show only fast moving objects, or to reduce  background noise.
    \item bias\_refr  adjusts the refractory period which determines the duration for which a pixel is blind after each event has been recorded. This can be used to change the number of events during a big illumination change without changing the sensitivity or the bandwidth. It is often used to make each big light change produce only one event at each pixel.
\end{enumerate}

In order to optimise the camera settings for our specific experimental scenario, we conducted a thorough evaluation of the most appropriate bias settings to attain the desired outcomes. Drawing insights from prior research in camera optimisation \cite{dilmaghani2023control}, we identified the bias\_hpf (high-pass filter) setting as the optimal choice for enhancing output quality for pulse detection. We set the value of this bias to 25 in the Metavision software which controls the camera as it provided  suitable picture quality while helping to reduce some of the background noise without having high latency or loss of events. 

\subsection{Event Detection}

A dot about the size of a 1 cent coin  drawn on the skin of the inside of the wrist using a  marker can create a high-contrast region compared to the surrounding skin. If the  dot is drawn directly over or near to where the radial artery pulsates on the wrist then the very slight motion on the surface of the skin as a result of pulsation which may not be visible to the human eye, will result in a periodic  change in light intensity in that region.  An event camera can be used to identify and record pixel-level events by detecting changes in lighting brought on by this contrast. As a result, the dot will appear to the event camera to have a periodically varying brightness when compared to the nearby skin.  
These variations in brightness produce events that stand out from the background  and give the event camera a distinct signal to recognise and record. 

\begin{figure}[ht]
    \centering
    \includegraphics[width=4.29cm]{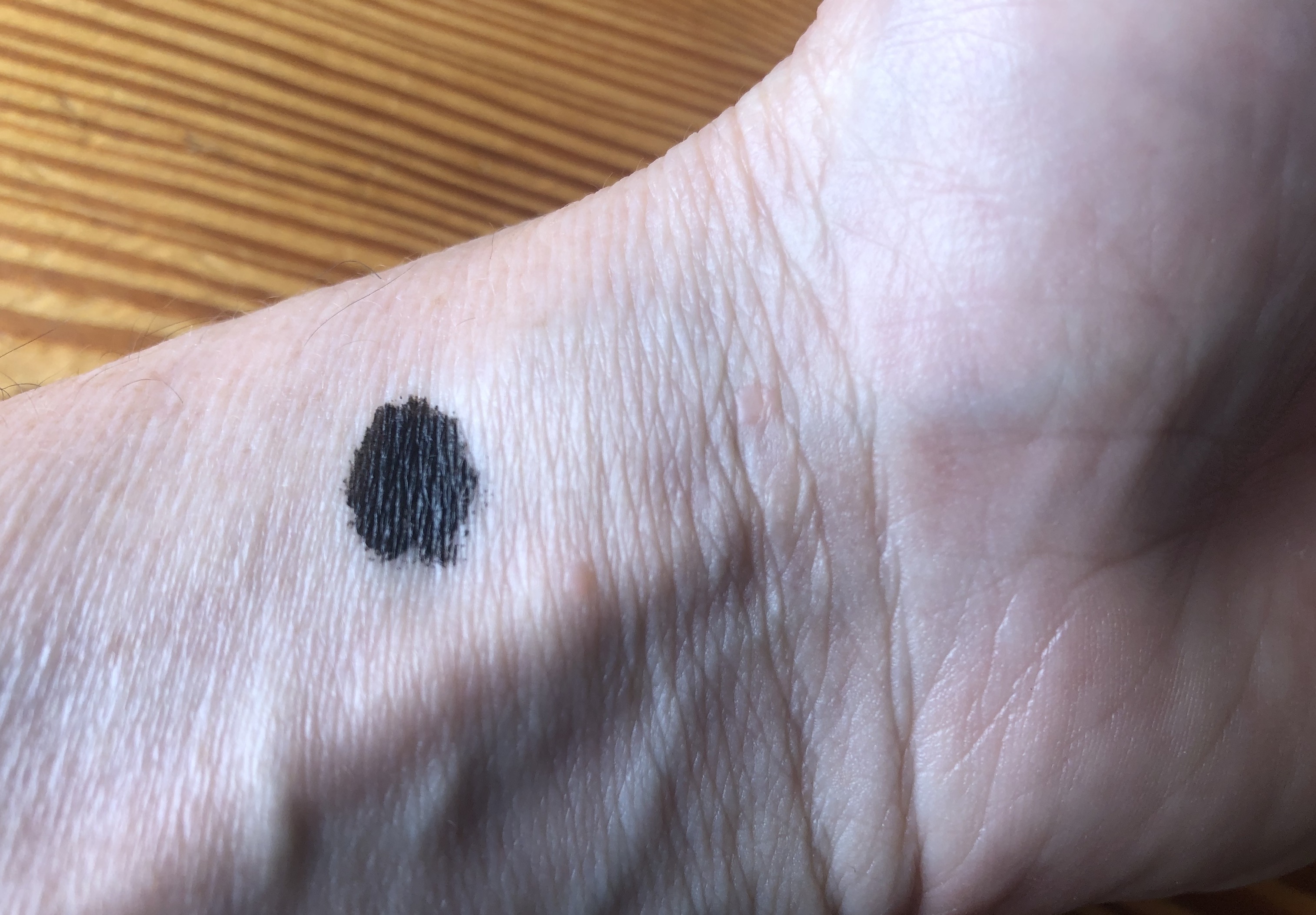}
    \includegraphics[width=4.29cm]{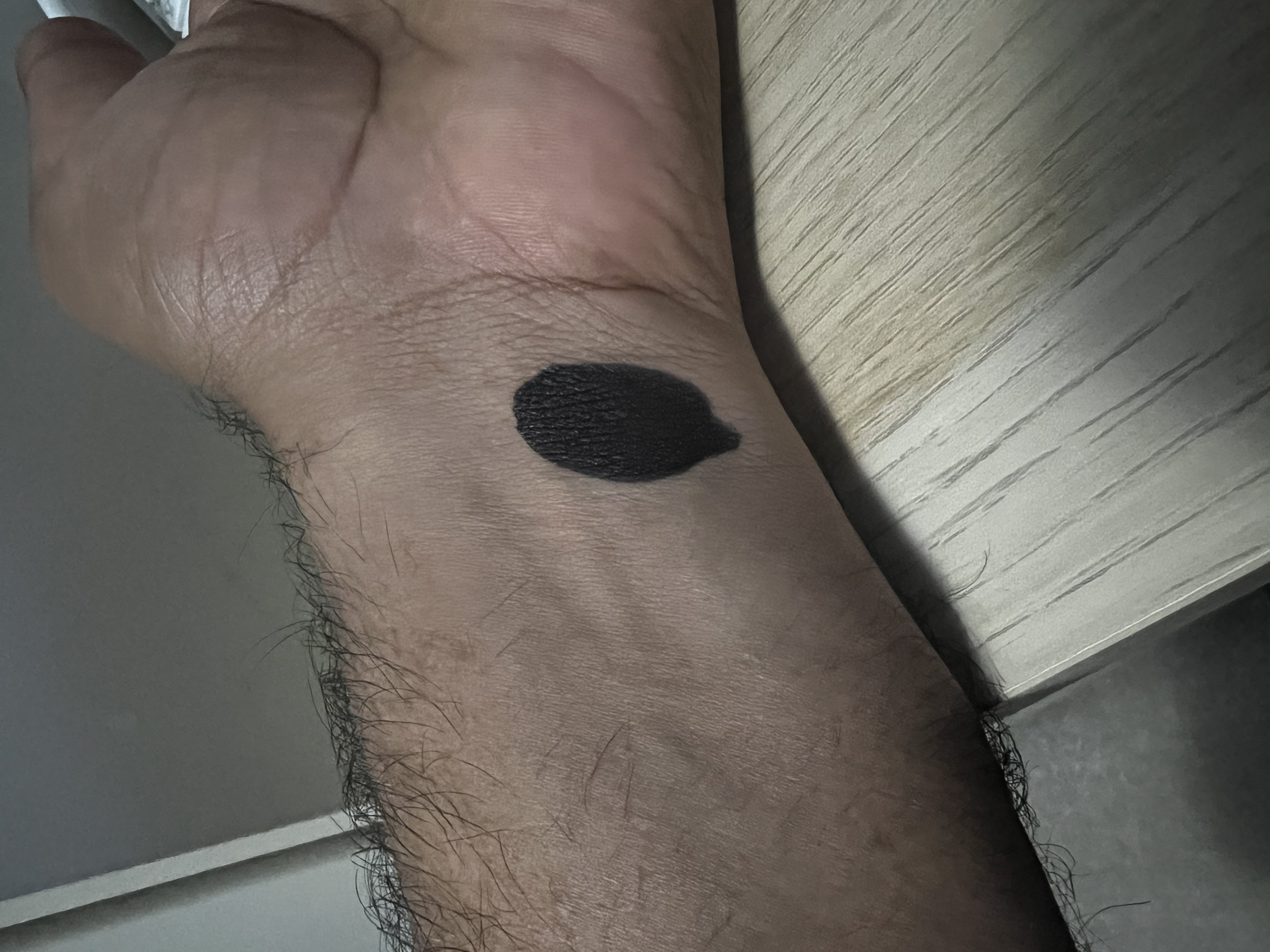}

        \vspace{0.15cm}
        
    \includegraphics[width=4.29cm]{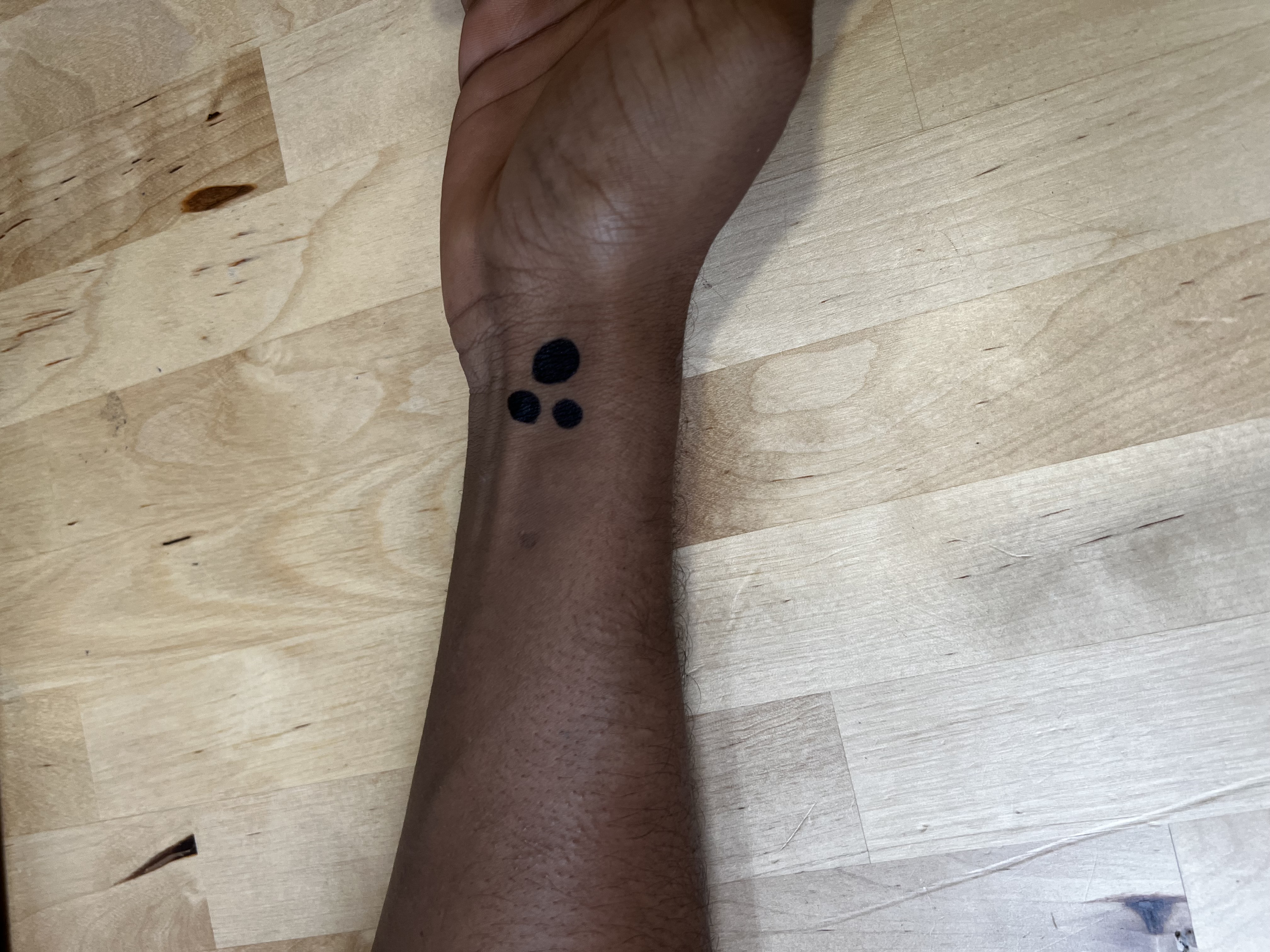}
    \includegraphics[width=4.29cm]{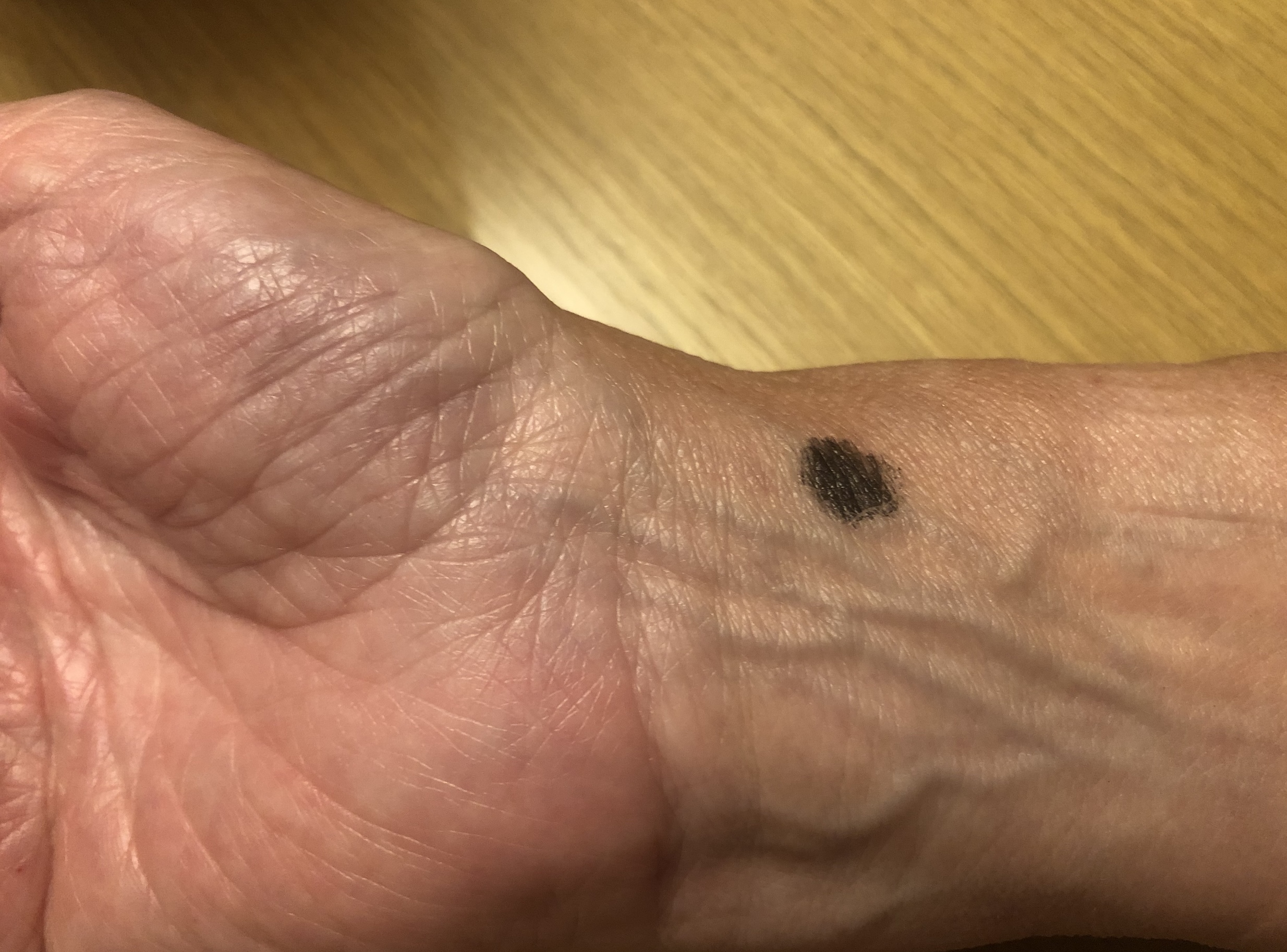}

        \caption{Samples of dots drawn on the wrists of some of our subjects
    \label{fig:wrists}}
\end{figure}

\begin{figure}[ht]
    \centering
       \includegraphics[width=0.5\textwidth]{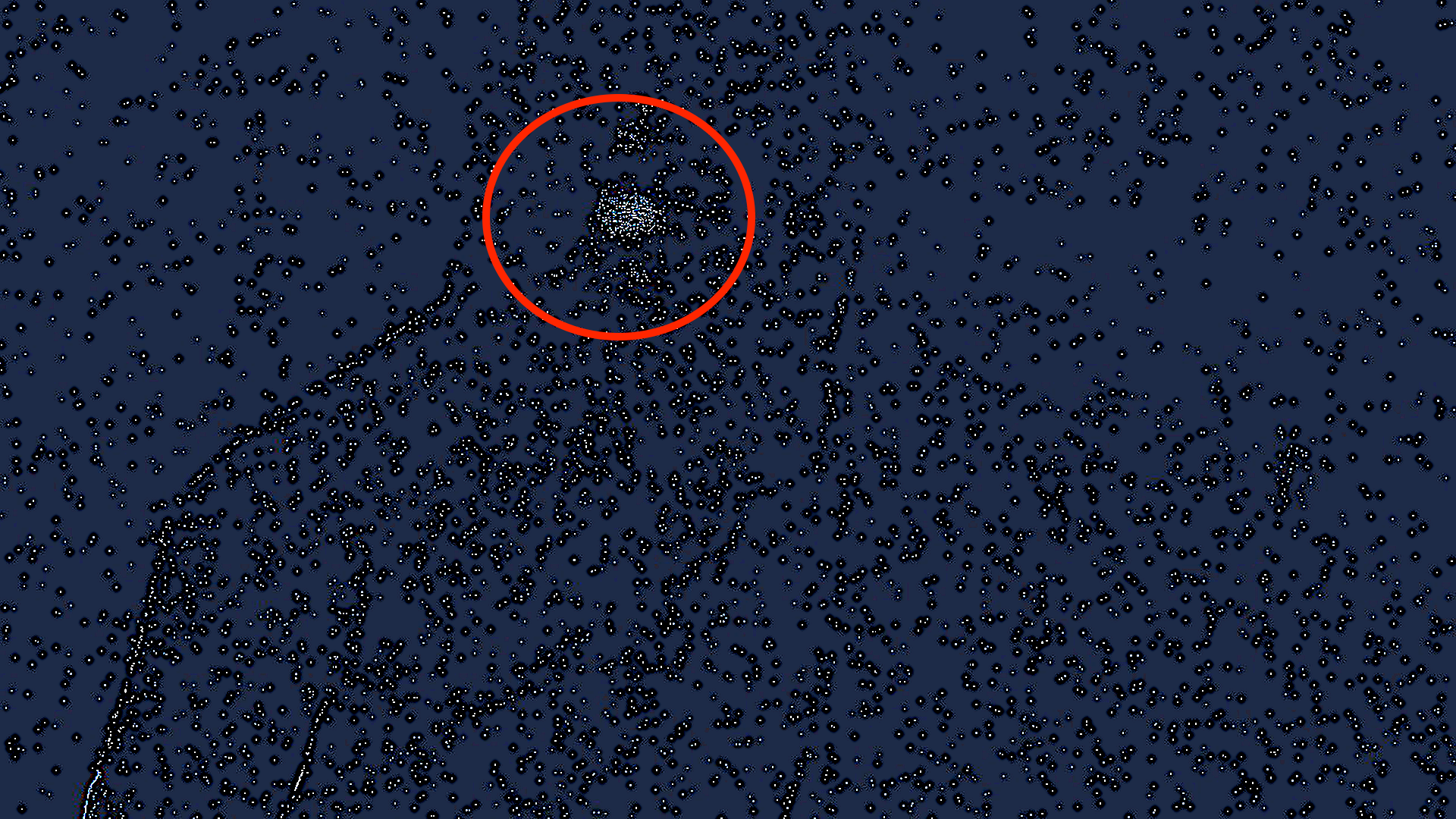}
        \caption{Screen-grab from Metavision Studio rendering of event camera recording of a subject's wrist. The placement of the coloured dot on the wrist can be seen in the top right part of the image as a cluster of white dots.
    \label{fig:metavis}}
\end{figure}

The designated dot may show up as a succession or a burst of events in the output stream, depending on  the  event capture settings in the camera, known as the bias settings.  Some sample dots drawn on the wrists of some of our subjects are shown in Figure~\ref{fig:wrists} while Figure~\ref{fig:metavis} shows a rendering of some of the pixel-level  events from an event stream  recording where each white dot corresponds to an event. In Figure~\ref{fig:metavis} there is a clear clustering of events from the black dots drawn on the wrist and highlighted by the red circle. The shape of the hand can be seen with the thumb to the lower left of the red circle. The line of events rendered as white dots and showing the outline of the hand corresponds to natural tremor, an always present and naturally occurring oscillatory motion in the human body while holding steady limb postures. This movement is frequent but not observable to the naked eye due to its very small amplitude \cite{marshall1956physiological} but is sufficient for the event camera to detect the change.

One of the main reasons event cameras can react differently when a dot is put on the skin with a  marker pen or when other rapid changes in the scene take place is their capacity to record such dynamic, high-frequency changes in illumination. This characteristic makes event cameras suitable for  applications where traditional cameras might not perform as effectively, such as in fast-paced dynamic environments or low-latency, high-frequency monitoring scenarios.

We recruited subjects for an experimental investigation into detecting heart rates from an event camera.  These were mostly University students and employees along with athletes attending a University gym. Their ages ranged from 20 to 50+ years and all were over 18 years of age. We  included  a variety of skin tones  and had an almost 50-50 male/female ratio among our subjects. 
Ethical approval for this work was granted by the School of Computing Research Ethics committee with subjects reading a plain language statement of their involvement and signing an informed consent form. 


\subsection{Setup and Data Acquisition}

\begin{figure}[!ht]
    \centering
    \includegraphics[width=0.3\textwidth]{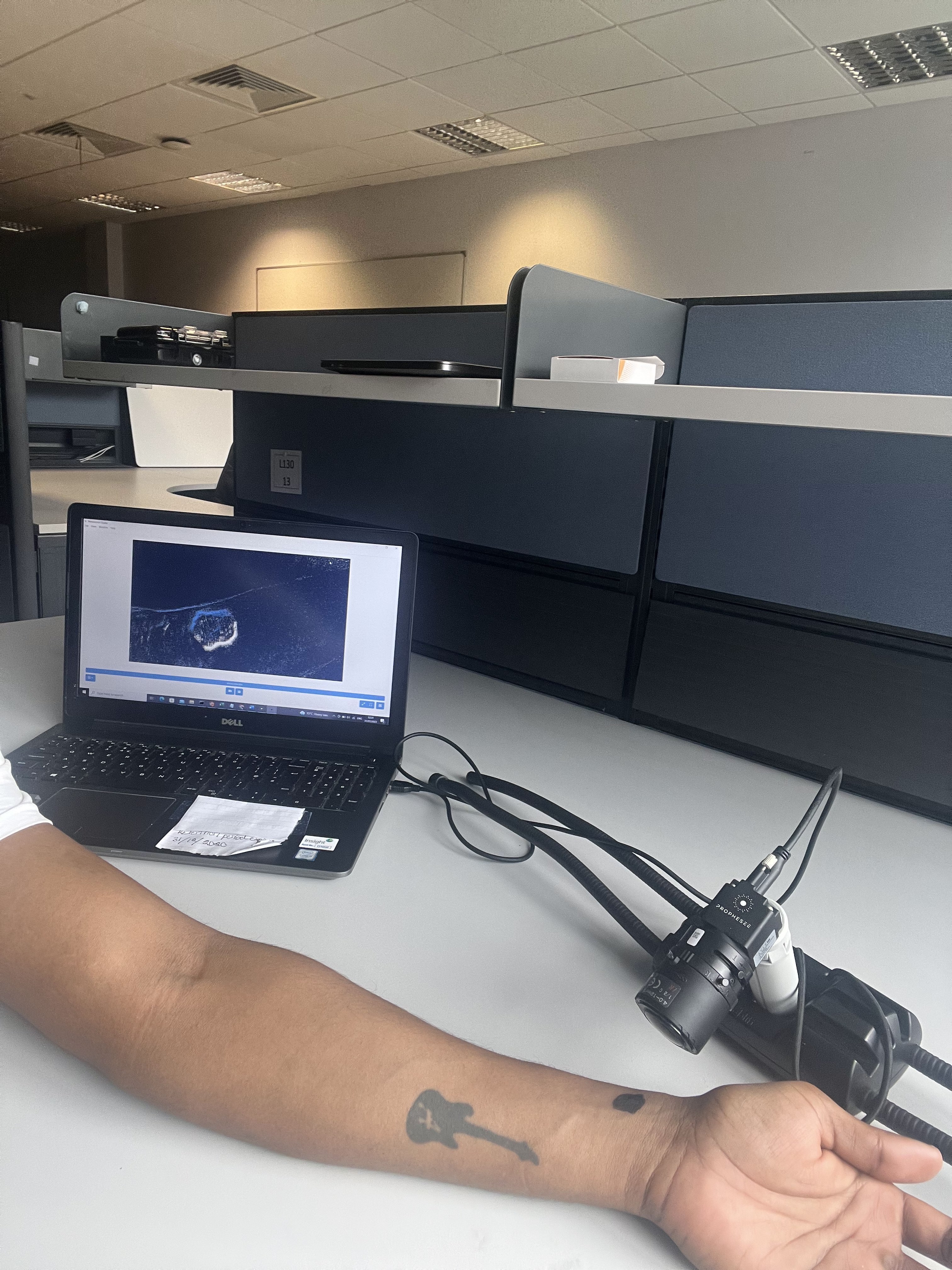}
        \caption{Laboratory setting for our data capture
    \label{fig:setup}}
\end{figure}

On arriving at our laboratory each subject wore an Apple watch to monitor their actual  heart rate  during the data capture, which was recorded manually.  We then marked a small dot on their wrist with a black marker, as shown  in Figure~\ref{fig:wrists}.
After the subject had relaxed and was rested and acclimatised to the laboratory environment, they were asked to place their arm on a desk where the event camera was set up as shown in Figure~\ref{fig:setup}. This is close to a window with natural daylight rather than use the flickering light, the subject was asked to remain still with their hand under the event camera for  12 to 15 seconds while a recording with the event camera was made  with the default bias settings on the event camera and a second recording was made with bias\_hpf (high-pass filter) value set to 25 in the Metavision software which controls the event camera, as described later.  At the same time that the event camera recordings were made  the heart rates as recorded by the Apple Watch they wore were  noted manually.

The subject was then asked to perform some form of indoor exercise of their choice to cause their heart rate to elevate. This could be an activity of their choice depending on their fitness levels and some did jumping jacks or star jumps for up to 1 minute. The elevated HR was monitored from the Apple watch  and recorded manually. The same procedure to capture two more recordings was repeated for the elevated HR.
The subject was then given a tissue and an alcohol-based sanitiser to remove the marked dot from their wrist.

\subsection{Data Overview}

In total we gathered data from 25 subjects as summarised  in Table~\ref{tab:results}. The dataset includes a range of subjects with different skin tones, age bands, fitness levels and heart rates. Some subjects had tattoos on their wrist which was useful to determine whether the sensor could still detect movement from pulsation. For each subject there are up to 4 event stream files recorded,  two  recorded at resting HR and  two recorded during elevated HR although 4 of our subjects declined to do exercises and to have an elevated heart rate reading giving us 46 event stream recordings  in total. The two files per heart rate per subject were recorded using the default bias setting and two more using a different bias setting. 

Each event stream file consists of a set of timestamps, x-y coordinates of pixels within the video frame and their polarity values (-1 or 1) depending on the change in brightness. 
For example, in one recording the pixel at the $(x,y)$ position $(346,142)$ in the frame had the following activations:
\begin{verbatim}
( 346, 142, 0, 235034 )
( 346, 142, 1, 237174 )
( 346, 142, 0, 238514 )
\end{verbatim}
meaning that at 0.235034 seconds the brightness decreased as indicated by the $0$, at 0.237174 seconds it then increased as indicated by the $1$ and 0.238514 seconds it decreased again. 
The number of events for each of these files may range from 850,000  to close to 3 million  depending on the duration of the recording and the amount of movement that may have occurred during that recording and how close the camera is placed to the wrist.  If the camera is close then the number of events is less while if it is further from the wrist then it captures more events because there are more edges to the hand and tremors detected.

\subsection{Deriving Heart Rate}

We now describe the series of steps we developed to calculate heart rates from the event stream camera data. 
\begin{enumerate}

    \item For each event stream file, we first calculate a heatmap for the  frame of size $1280 \times 720$ which is the resolution of the Prophesee EVK4 camera, corresponding to the number of all event activations at each $x,y$ pixel co-ordinate.
    
    \item We then identify the $100 \times 100$ pixel area within the frame for each event stream where the sum of all pixel activations is highest, which we refer to as our area of interest (AoI). 
    
    \item We divide the detected AoI into smaller, nonoverlapping tiles of size $5 \times 5$ pixels and we quantise all of the events in the event stream which occurred for each of these tiles. The individual events are timestamped to the nearest 0.001ms and we divide the events in each tile into bins or ranges of 1/50 seconds duration. 

    \item For each of the $5 \times 5$ pixel regions we determine the dominant frequency from a periodogram which is an estimate of power spectral density (PSD) which is described earlier in Section~\ref{sec:periodogram} and which is implemented in Python.\footnote{We used the Python funciton scipy.signal.periodogram  from the essential signal processing package scipy \cite{scipy}.}  We then fuse those dominant periodicity frequencies to give the  estimated pulse rate for the recording. The PSD describes  is calculated using the Fast Fourier Transform (FFT) algorithm.

\end{enumerate}

During data capture and data processing we  encountered a number of operational challenges as follows. The first was that  subjects need to maintain a steady hand position and this was not always the case because of natural tremor \cite{marshall1956physiological} and because some subjects did not relax completely and were tense from holding their arm in an unnatural position.  We also had to ensure that  lighting conditions were constant by recording  near to a window in order to use natural light rather than use controlled lighting. This has the advantage of more accurately replicating a real world use case but meant that lighting intensity varied across  recordings depending on the level of sunlight at recording time.

\section{Experimental Results}

We gathered heart rate data and event camera recordings from 25 subjects but 4 of these  declined to do exercise in the lab setting to elevate their heart rates so we had a total of 46 pulse rates from our 25 test subjects, with 4 event stream recordings for most of our subjects. After running our pulse detection algorithm our results are presented in Table~\ref{tab:results} and show that we were able to detect pulse rates for 40 of the 46 recordings. For the other 6 recordings we found that there had been an excess of the naturally-occurring sub-conscious movements or tremors in the hand for 3 of the recordings and for 3 others we have not been able to pinpoint the root cause for nondetection.  
\begin{table*}[!htb]
 \caption{Characteristics of 25 subjects, their actual resting and elevated heart rates (HRs) and the HRs as determined by our event camera algorithm where ``ND" corresponds to not detected by our algorithm and ``-" corresponds to a subject declining the invitation to provide data for an elevated HR. The column headed ``Biases Used'' indicates which camera setting, either the default (D) or customised high-pass filter (C) gave the best result or if both showed a similar result (B).\label{tab:results}  } 
 \centering
\begin{tabular}{ccccccc}
\toprule
\multirow{2}{*}{Age} & \multirow{2}{*}{Gender} & Skin   & Resting/Active  & Resting/Active  & \multirow{2}{*}{Differences} & Biases\\
&&Tone& \multicolumn{1}{c}{Actual HR} & \multicolumn{1}{c}{Detected HR} && Used \\
\midrule
20-30 & M & Brown  & 66 / 93 & 65 / 91 & -1 / -2 & C / D \\
20-30 & F & Brown  & 66 / 111 & 65 / 114 & -1 / +3   & C / B \\
50-60 & F & White  & 78 / - & 79 / -  & +1 / -  & C / - \\
20-30 & M & White  & 80 / 116 & 80 / 116 & 0 / 0   & D / B \\
20-30 & M & Black  & 63 / 118 & 63 / ND & 0 / ND   & B / - \\
20-30 & F & White  & 69 / - & 69 / - & 0 / -   & B / - \\
30-40 & M & Black  & 64 / 128 & 65 / 129 & +1 /+1  & D / B \\
20-30 & F & White  & 81 / 105 & 79 / 105 & -2 / 0   & C / B \\
20-30 & F & White  & 72 / - & 71 / - & -1 / -   & D / - \\
30-40 & M & Brown  & 58 / 95 & 58 / 96 & 0 / +1  & B / D \\
20-30 & M & White  & 93 / 98 & 94 / 98 & +1 / 0   & B / B \\
20-30 & M & White  & 57 / 86 & 58 / 83 & +1 / -3  & B / D \\
20-30 & F & Black  & 76 / 102 & ND / 105 & ND / +3  & - / C \\
20-30 & F & White  & 79 / 113 & 76 / 113 & -3 / 0   & C / D \\
20-30 & F & Brown &  80 / 104  & 81 / 103 & +1 /-1  & B / C \\
20-30 & F & White &  84 / - & 86 / - & +2 / -   & D / - \\
30-40 & F & Black &  67 / 129 & 64 / 126 & -3 / -3   & D / C \\
20-30 & M & White &  71 / 105 & 71 / 101 & 0 / -4  & C / B \\
20-30 & F & White &  96 / 154 & ND / ND  & ND / ND  & - / - \\
20-30 & M & Brown &  107 / 115 & 102 / 114 & -5 / -1   & D / C \\
20-30 & M & White &  73 / 102 & 75 / 103 & +2 / +1  & D / C \\
20-30 & M & Black &  79 / 132 & 78 / ND & -1 / ND  & C / - \\
20-30 & F & Brown &  109 / 130 & 109 / 129 & 0 / -1   & C / C \\  
20-30 & F & White &  67 / 161 & 63 / 156 & -4 / -5  & C / C \\
20-30 & M & Brown &  70 / 101 & 66 / ND & -4 / ND   & C / - \\ 
\bottomrule
 \end{tabular}
\end{table*}

The mean absolute error (MAE) and root mean squared error (RMSE) values for our estimates of resting and elevated HRs from 23 and from 17 subjects respectively are shown in Figure~\ref{tab:performance_summary}.
\begin{table}[htb]
    \centering
        \caption{MAE and RSME for resting HR (23) and elevated HR (17 subjects)}
    \label{tab:performance_summary}
    \begin{tabular}{lcc}
    \toprule
    & MAE & RMSE \\
    \midrule
  Resting HR~~~~~~~~~ & 1.478  & 2.043 \\
  Elevated HR & 1.706 & 2.262 \\
  \bottomrule
    \end{tabular}
\end{table}

For the heart rates that we did detect, the highest difference between the actual  and  detected pulse was 5 beats per minute (bpm) occurring just twice, and  in  24 of the 40 cases, the pulse was detected precisely or within 1~bpm. 
The mean absolute error (MAE) and root mean squared error (RMSE) values for our estimates of resting and elevated HR from 23 and from 17 subjects respectively are shown in Table~\ref{tab:performance_summary} and these are less than 2~bpm for MAE and just over 2~bpm for RMSE. This compares favourably with the MAE/RMSE figures of 2.11/2.93, 2.43/3.44, and 2.26/3.45 bpm for biking, stepping, and treadmill exercises respectively, as reported in \cite{9654212} and which is based on noncontact photoplethysmography.
The best-performing camera bias settings was the customised high pass filter giving the best or joint-best performance on 29 of the 40 recordings.

\begin{figure*}[!ht] 
    \begin{center} 
        \includegraphics[width=0.7\textwidth]{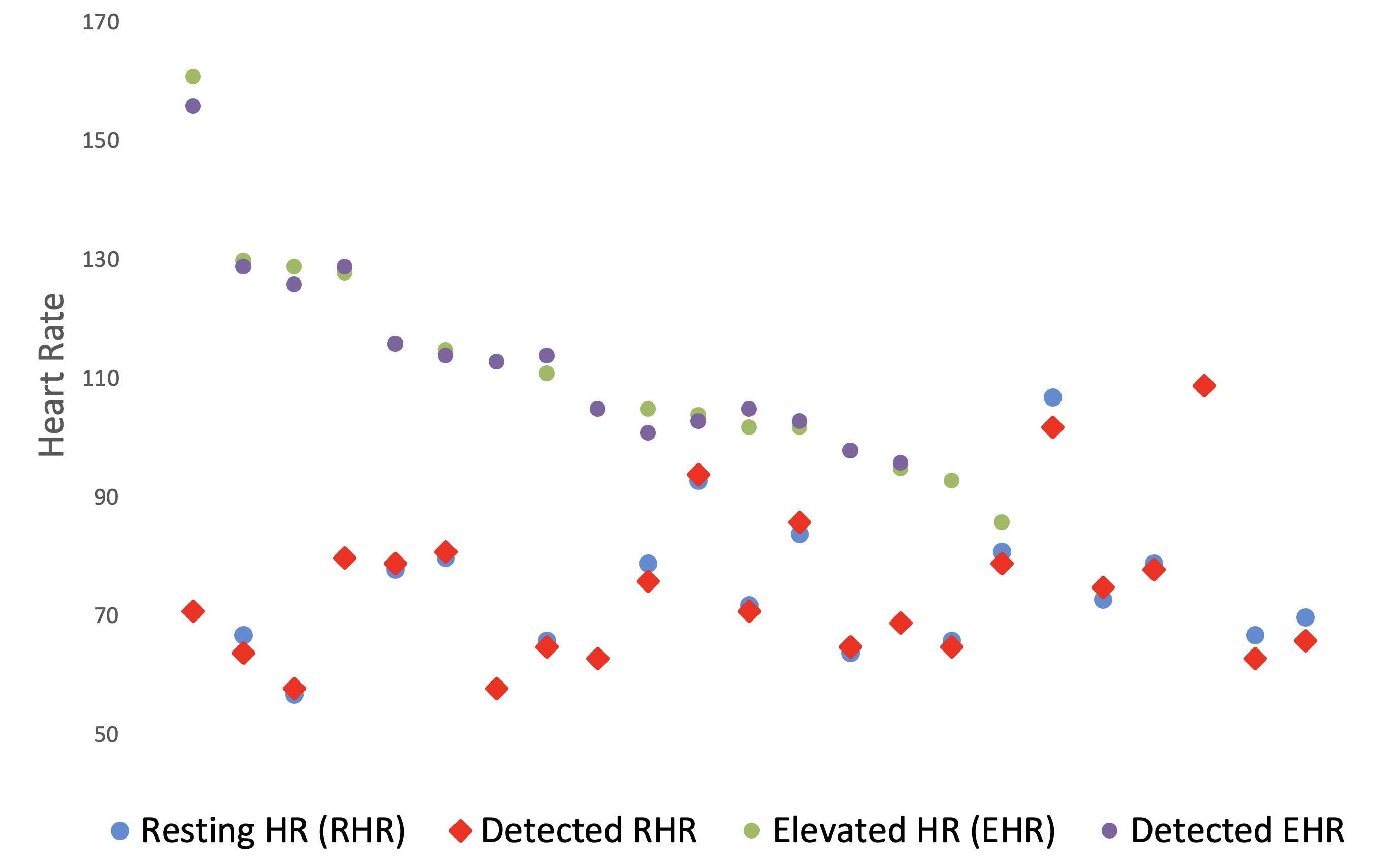}
    \end{center}    
    \caption{Actual vs. estimated HRs for both the elevated and resting HR settings. Where actual and estimated are exactly the same or very close one graph marker occludes the other.
    \label{fig:results}}
\end{figure*}

Figure~\ref{fig:results} shows a graph of the actual vs. estimated HRs for both resting and elevated settings for each subject  with subjects sorted by the value of decreasing elevated HR.  The differences between the  pairs of points, the actual vs. estimated HRs, reflects the accuracy of our estimations.
The variation between actual and estimated HRs could also be attributed to the fact that there was a slight delay between the subjects doing exercise and the recording of their heart rates using the Apple watch.

Using an event camera to detect a person's pulse from their wrist is a new idea with potential use by the automotive industry for driver monitoring and driver safety, for example. Our results show that the concept of using an event camera for detecting physiological signals, specifically pulse rate, is feasible.

\section{Conclusions and Further Work}

In this paper we have used an event camera for noncontact estimation of heart rate from the wrists of 25 subjects where each subject had a black mark written onto their skin.  After identifying the region of the event stream frame corresponding to where the black dot is located, we identified the dominant frequency from a periodicity analysis of event camera events occurring at that region of the frame and that becomes our estimate for the subject's heart rate at that time.  

We applied this technique to event camera data from the wrists of 46 recordings of 25 subjects of diverse ages and skin colours and our results compare favourably with other noncontact estimation techniques based on photoplethysmography. Our technique achieves 1.478/2.043 and 1.706/2.262 (MAE/RMSE) for resting and for elevated HR respectively which is comparable to the  2.11/2.93, 2.43/3.44, and 2.26/3.45 bpm for biking, stepping, and treadmill exercises, respectively  when compared to a commercial Polar H7 chest strap as reported in \cite{9654212}.
In summary, when examining the  results, we can say that HR detection with event cameras is not only possible but has been demonstrated, though it does need further research to improve its applicability.

While our investigation into the potential for using event-based cameras for HR detection has demonstrated it is possible to an acceptable leverl of accuracy, several avenues of further research can be explored to enhance the robustness and applicability of this approach.

Developing a real-time algorithm for pulse detection using event cameras is crucial for their practical use in applications like remote patient monitoring, driver awareness monitoring and fitness tracking. Optimising the computational efficiency of the algorithms while maintaining accuracy would be a significant focus for further research as would operating in variable and uncontrolled lighting conditions and catering for subject movement during monitoring.
Research  should focus on determining pulse while the camera device is pointed at the subject rather than ex-ante computation.

Finally, as with any remote monitoring technology, privacy and ethical concerns also need to be addressed.  
Conducting field studies to assess the practical usability and user experience of HR detection based on event cameras  in real-world scenarios is also important. Understanding user acceptance, comfort, and satisfaction would be crucial for widespread adoption and use of the technology.

\vspace{0.5cm}
\noindent 
{\bf Note:} All of the data used in the experiments in this paper has been made publicly available at \url{https://doi.org/10.6084/m9.figshare.24039501.v1}.

\bibliographystyle{plain} 
\bibliography{ISM2023.bib} 

\begin{thebibliography}{10}

\bibitem{BERES2021102589}
Szabolcs Béres and László Hejjel.
\newblock The minimal sampling frequency of the photoplethysmogram for accurate
  pulse rate variability parameters in healthy volunteers.
\newblock {\em Biomedical Signal Processing and Control}, 68:102589, 2021.

\bibitem{chen2020event}
Guang Chen, Hu~Cao, Jorg Conradt, Huajin Tang, Florian Rohrbein, and Alois
  Knoll.
\newblock Event-based neuromorphic vision for autonomous driving: A paradigm
  shift for bio-inspired visual sensing and perception.
\newblock {\em IEEE Signal Processing Magazine}, 37(4):34--49, 2020.

\bibitem{9654212}
Yung-Chien Chou, Bo-Yi Ye, Hong-Ren Chen, and Yuan-Hsiang Lin.
\newblock A real-time and non-contact pulse rate measurement system on fitness
  equipment.
\newblock {\em IEEE Transactions on Instrumentation and Measurement}, 71:1--11,
  2022.

\bibitem{scipy}
The~SciPy community.
\newblock Estimate power spectral density using a periodogram.
\newblock Available at
  \url{https://docs.scipy.org/doc/scipy/reference/generated/scipy.signal.periodogram.html},
  2023.
\newblock Last accessed 24 August, 2023.

\bibitem{dilmaghani2023control}
Mehdi~Sefidgar Dilmaghani, Waseem Shariff, Cian Ryan, Joe Lemley, and Peter
  Corcoran.
\newblock Control and evaluation of event cameras output sharpness via bias.
\newblock In {\em Fifteenth International Conference on Machine Vision (ICMV
  2022)}, volume 12701, pages 455--462. SPIE, 2023.

\bibitem{gallego_event-based_2022}
Guillermo Gallego, Tobi Delbrück, Garrick Orchard, Chiara Bartolozzi, Brian
  Taba, Andrea Censi, Stefan Leutenegger, Andrew~J. Davison, Jörg Conradt,
  Kostas Daniilidis, and Davide Scaramuzza.
\newblock Event-{Based} {Vision}: {A} {Survey}.
\newblock {\em IEEE Transactions on Pattern Analysis and Machine Intelligence},
  44(1):154--180, January 2022.

\bibitem{holesovsky_experimental_2021}
Ondrej Holesovsky, Radoslav Skoviera, Vaclav Hlavac, and Roman Vitek.
\newblock Experimental {Comparison} between {Event} and {Global} {Shutter}
  {Cameras}.
\newblock {\em Sensors}, 21(4):1137, February 2021.

\bibitem{kielty2023neuromorphic}
Paul Kielty, Cian Ryan, Mehdi~Sefidgar Dilmaghani, Waseem Shariff, Joe Lemley,
  and Peter Corcoran.
\newblock Neuromorphic seatbelt state detection for in-cabin monitoring with
  event cameras.
\newblock {\em arXiv preprint arXiv:2308.07802}, 2023.

\bibitem{s17071490}
Yu-Chen Lin, Nai-Kuan Chou, Guan-You Lin, Meng-Han Li, and Yuan-Hsiang Lin.
\newblock A real-time contactless pulse rate and motion status monitoring
  system based on complexion tracking.
\newblock {\em Sensors}, 17(7), 2017.

\bibitem{marshall1956physiological}
John Marshall and E~Geoffrey Walsh.
\newblock Physiological {T}remor.
\newblock {\em Journal of Neurology, Neurosurgery, and Psychiatry}, 19(4):260,
  1956.

\bibitem{prophesee_doccumentation}
{PROPHESEE Metavision For Machines}.
\newblock {Biases}-{Metavision SDK Docs 4.2.1 documentation}.
\newblock Available
  at:\url{https://docs.prophesee.ai/stable/hw/manuals/biases.html}.
\newblock [Accessed 26-07-2023].

\bibitem{rong2021radar}
Yu~Rong, Kumar~Vijay Mishra, and Daniel~W Bliss.
\newblock Radar-based radial arterial pulse rate and pulse pressure analysis.
\newblock In {\em 2021 29th European Signal Processing Conference (EUSIPCO)},
  pages 1870--1874. IEEE, 2021.

\bibitem{2023}
Alan~F. Smeaton and Feiyan Hu.
\newblock Periodicity intensity reveals insights into time series data: Three
  use cases.
\newblock {\em Algorithms}, 16(2):119, Feb 2023.

\end{thebibliography}


\end{document}